\documentclass[usenatbib]{mnras}

\usepackage{graphicx}% Include figure files
\usepackage{dcolumn}% Align table columns on decimal point
\usepackage{amssymb} %maths
\usepackage{ctable}
\usepackage{bm}% bold math
\usepackage{epsfig}
\usepackage{epstopdf}
\usepackage{epsf,color}
\usepackage{aas_macros}
\usepackage{url}
\usepackage{widetext}

\newcommand{\cmnt}[1]{}

\newcommand{\VEV}[1]{\left\langle #1 \right\rangle}

\title[Constraining Gravity through CMB Lensing]{Constraining Gravity at the Largest Scales through CMB Lensing and Galaxy Velocities}

\author[A.~Pullen et al.]{Anthony R. Pullen\thanks{Email: apullen@andrew.cmu.edu}, Shadab Alam, Siyu He, and Shirley Ho\\ McWilliams Center for Cosmology, Department of Physics, Carnegie Mellon University, 5000 Forbes Ave, Pittsburgh, PA, 15213, U.S.A.}

\date{Accepted XXX. Received YYY; in original form ZZZ}

\pubyear{2016}

\begin{document}
\label{firstpage}
\pagerange{\pageref{firstpage}--\pageref{lastpage}}
\maketitle

\begin{abstract}
We demonstrate a new method to constrain gravity on the largest cosmological scales by combining measurements of cosmic microwave background (CMB) lensing and the galaxy velocity field.  $E_G$ is a statistic, constructed from a gravitational lensing tracer and a measure of velocities such as redshift-space distortions (RSD), that can discriminate between gravity models while being independent of clustering bias and $\sigma_8$. While traditionally, the lensing field for $E_G$ has been probed through galaxy lensing, CMB lensing has been proposed as a more robust tracer of the lensing field for $E_G$ at higher redshifts while avoiding intrinsic alignments.  We perform the largest-scale measurement of $E_G$ ever, up to 150 Mpc/$h$, by cross-correlating the \emph{Planck} CMB lensing map with the Sloan Digital Sky Survey III (SDSS-III) CMASS galaxy sample and combining this with our measurement of the CMASS auto-power spectrum and the RSD parameter $\beta$.  We report $E_G(z=0.57)=0.243\pm0.060\,{\rm (stat)}\pm0.013\,{\rm (sys)}$, a measurement in tension with the general relativity (GR) prediction at a level of 2.6$\sigma$.  Note that our $E_G$ measurement deviates from GR only at scales greater than 80 Mpc/$h$, scales which have not been probed by previous $E_G$ tests.  Upcoming surveys, which will provide an order-of-magnitude reduction in statistical errors, can significantly constrain alternative gravity models when combined with better control of systematics.
\end{abstract}
%\maketitle

\section{Introduction}

Since cosmic acceleration was first discovered \citep{1998AJ....116.1009R,1999ApJ...517..565P}, there have been many investigations seeking to determine its cause.  The cosmological constant, a form of dark energy \citep{2003RvMP...75..559P} that exhibits a negative pressure $p=-\rho$, can explain the cosmic acceleration and is consistent with measurements of the cosmic microwave background (CMB) \citep{2013ApJS..208...20B,2015arXiv150201589P,2015arXiv150201590P} and large-scale structure (LSS) \citep{2014MNRAS.441...24A}.  However, if gravity were weaker than predicted by general relativity (GR) on cosmological scales, then this could also cause the cosmic expansion to accelerate.  This concept, called \emph{modified gravity}, cannot be distinguished from dark energy by measuring the cosmic expansion, \emph{i.e.} through supernova \citep{2011ApJ...737..102S} or baryon acoustic oscillations \citep{2014MNRAS.441...24A}, alone, requiring a measurement of the growth of structure, \emph{e.g.} through redshift-space distortions (RSD) \citep{1987MNRAS.227....1K,1998ASSL..231..185H,2012MNRAS.423.3430B,2014MNRAS.439.3504S,2015MNRAS.453.1754A,2016MNRAS.456.3743A}, to break this degeneracy.  Several upcoming observatories hope to test general relativity on cosmological length scales using these methods.

$E_G$ \citep{2007PhRvL..99n1302Z} is a statistic that probes gravity by measuring the ratio between curvature and velocity perturbations using measurements of gravitational lensing, galaxy clustering, and growth of structure.  This quantity is a member of a general class of parametrized deviations from GR [\emph{e.g.} see \citet{2011JCAP...08..005H}, \citet{2016PhRvD..93b3513D}, and references therein].  $E_G$ is related to the Poisson field equation which is modified between various gravity models, breaking the degeneracy and model dependence in current cosmological probes of gravity and dark energy.  It is also independent of clustering bias on linear scales; thus unlike probes of gravity using measurements of structure growth directly, the clustering bias does not have to be modeled or marginalized in $E_G$ measurements.  The lensing signal within $E_G$ has traditionally been probed with galaxy-galaxy lensing \citep{2013MNRAS.432.1544M}, or lensing of galaxies by foreground galaxies.  In \citet{2010Natur.464..256R}, $E_G$ was measured at $z=0.32$ over scales 10--50 Mpc/$h$ to be 0.39$\pm$0.06.  Recently, a measurement of $E_G$ from galaxy lensing was performed using several datasets \citep{2016MNRAS.456.2806B}, finding, over scales 10--50 Mpc/$h$, to be 0.48$\pm$0.10 at $z=0.32$ and 0.30$\pm$0.07 at $z=0.57$.  All these measurements are consistent within 1$\sigma$ with predicted GR values.  In other work, constraints for future galaxy lensing surveys were forecasted \citep{2015JCAP...12..051L}.

It has recently been proposed \citep{2015MNRAS.449.4326P} to measure $E_G$ using galaxy-CMB lensing \citep{2015arXiv150201591P}, a more robust lensing tracer that can probe $E_G$ at earlier times and larger scales than is currently possible with galaxy lensing. In addition, measuring $E_G$ using CMB lensing has advantages over galaxy-galaxy lensing.  Source galaxies in galaxy-galaxy lensing are usually assigned photometric redshifts with non-negligible uncertainties and can only be lensed by foreground galaxies at low redshifts.  For CMB lensing, the source redshift, $z=1100$ is well known and extremely high relative to galaxies ($z\sim 1$), allowing probes of $E_G$ at much higher redshifts.  Also, the intrinsic distribution of CMB photons is nearly Gaussian \citep{2015arXiv150201592P} and is not affected by complex astrophysical effects, such as intrinsic alignments in galaxy lensing.  It was shown in \citet{2015MNRAS.449.4326P} that Advanced ACTPol \citep{2016JLTP..tmp..144H}, an upcoming CMB survey, combined with a spectroscopic galaxy survey, \emph{e.g.} the Dark Energy Spectroscopic Instrument (DESI) \citep{2013arXiv1308.0847L}, would measure $E_G$ at 2\% precision, or $<1$\% precision with a photometric survey, \emph{e.g.} the Large Synoptic Survey Telescope (LSST) \citep{2009arXiv0912.0201L}.  Recently, \citet{2016MNRAS.456.3213G} proposed a bias-independent statistic $D_G$, an alternative to $E_G$ that does not include growth information.  This work was also able to measure $D_G$ using the \emph{Planck} CMB lensing map and the Dark Energy Survey (DES) \citep{2005astro.ph.10346T} with photometric redshifts.  However, unlike $E_G$, $D_G$ cannot be directly related to modified gravity models.

In this analysis, we measure $E_G$ by combining measurements of the CMB lensing convergence map \citep{2015arXiv150201591P} from the latest \emph{Planck} data release \citep{2015arXiv150201582P} with the galaxy distribution from the CMASS galaxy sample \citep{2014MNRAS.441...24A} from the Sloan Digital Sky Survey (SDSS) III \citep{2011AJ....142...72E}.  We also test for various systematic effects in both the CMASS galaxy map and the \emph{Planck} CMB lensing convergence map.  We find $E_G(z=0.57)=0.243\pm0.060\,{\rm (stat)}\pm0.013\,{\rm (sys)}$, which is in tension with the expected $\Lambda$CDM value of $E_G(z=0.57|{\rm GR})=0.402\pm0.012$.  This tension appears at scales greater than 80 Mpc/$h$, scales which have not been probed by previous $E_G$ measurements.  By probing gravity over the scales 23--150 Mpc/$h$, this is the largest-scale measurement of $E_G$ ever performed, and only next-generation surveys, \emph{e.g.}~\emph{Euclid} \citep{2011arXiv1110.3193L}, will be able to probe these scales with $E_G$ using galaxy lensing.

The plan of our paper is as follows: in Section \ref{S:eg} we review the expression for $E_G$ and how we estimate it, and we describe the \emph{Planck} and CMASS data products we use in Section \ref{S:data}.  In Section \ref{S:cl} we describe how our angular power spectrum models are constructed.  We describe our estimators for the angular power spectra and $\beta$ in Section \ref{S:estim}, and we present our results in Section \ref{S:results} and estimates of systematic errors in Section \ref{S:cmbgalsys}.  We conclude in Section \ref{S:conclude}.

\section{$E_G$ Formalism} \label{S:eg}
Here we present a brief review of the expression for $E_G$ and how it is measured.  For a more comprehensive presentation, see \citet{2007PhRvL..99n1302Z} and \citet{2015MNRAS.449.4326P}.

The quantity $E_G$ is given by the expression (in Fourier space)
\begin{eqnarray}\label{E:egthk}
E_G(k)=\frac{c^2k^2(\phi-\psi)}{3H_0^2(1+z)\theta(k)}\, ,
\end{eqnarray}
assuming a flat universe described by a Friedmann-Robertson-Walker (FRW) metric, where $H_0$ is Hubble's constant, $\theta(k)$ is the perturbation in the divergence of the velocity field, and $\psi$ and $\phi$ are the time and space perturbations in the FRW metric.  On linear scales, $\theta(k)=f(z)\delta(k)$, where $\delta$ is the matter field perturbation, and $f(z)$ is the logarithmic rate of structure growth, also known as the growth rate.  By assuming GR, non-relativistic matter species, and no anisotropic stress, it can be shown using the Poisson equation from GR that Eq.~\ref{E:egthk} simplifies to
\begin{eqnarray}
E_G=\frac{\Omega_{m,0}}{f(z)}\, ,
\end{eqnarray}
where $\Omega_{m,0}$ is the relative matter density today and $f(z)\simeq[\Omega_m(z)]^{0.55}$ is the growth rate for GR.  Note that $E_G$ for $\Lambda CDM$ with GR is scale-independent.  For modified gravity theories, the expressions for $E_G(z)$ and $f(z)$ can be altered, producing values for $E_G$ that are distinct from GR and possibly scale-dependent.

$E_G$ can be estimated as
\begin{eqnarray} \label{E:egth}
E_G(\ell)=\Gamma\frac{C_\ell^{\kappa g}}{\beta C_\ell^{gg}}\, ,
\end{eqnarray}
where $\Gamma$ is a prefactor depending on Hubble parameter $H(z)$, the CMB lensing kernel $W(z)$, and the galaxy redshift distribution $f_g(z)$ (see Eq.~\ref{E:gamma}), $C_\ell^{\kappa g}$ is the CMB lensing convergence-galaxy angular cross-power spectrum, $C_\ell^{gg}$ is the galaxy angular auto-power spectrum, and $\beta$ is the redshift space distortion parameter.  In the linear perturbation regime, $\beta=f/b_g$ where $f$ is the growth rate and $b_g$ is the clustering bias of galaxies relative to matter perturbations.  Note that $\kappa$ is the lensing convergence, which is a line-of-sight integral of $\nabla^2(\psi-\phi)$ over the lensing kernel.  As in previous measurements using galaxy-galaxy lensing, $E_G$ measured using CMB lensing is independent of clustering bias and the amplitude of matter perturbations parametrized by $\sigma_8$, eliminating the need for measurements of (or marginalizing over) $b_g$ and $\sigma_8$ as in other gravity probes.

\section{Data} \label{S:data}
\subsection{Cosmic Microwave Background Lensing Map}

The cosmic microwave background (CMB) lensing map was provided by the Planck Collaboration \citep{2015arXiv150201582P}.  The \emph{Planck} satellite observed the intensity and polarization fields of the cosmic background radiation (CBR) over the whole sky.  The CBR was measured between August 2009 and August 2013 using an array of 74 detectors  consisting of two instruments.  The Low-Frequency Instrument (LFI) \citep{2010A&A...520A...4B,2011A&A...536A...3M} consists of pseudo-correlation radiometers and contains 3 channels with frequencies 30, 40 and 70 GHz.  The High-Frequency Instrument (HFI) \citep{2010A&A...520A...9L,2011A&A...536A...6P} consists of bolometers and contains 6 channels with frequencies 100, 143, 217, 353, 545, and 857 GHz.  These maps were combined and foreground-cleaned using the SMICA code \citep{2015arXiv150205956P}  to produce temperature and $E$ and $B$ polarization maps of the CMB with HEALPix \citep{2005ApJ...622..759G} pixelization with $N_{\rm side}=2048$ over approximately 70\% of the sky.  The temperature and polarization maps over all available frequencies are combined to reconstruct the minimum-variance CMB lensing field \citep{2015arXiv150201591P}; however, most of the lensing information comes from the 143 GHz and 217 GHz maps.  These channels have Gaussian beams with full-width-at-half-maxima (FWHMs) of 7' and 5', respectively, and temperature (polarization) noise levels of 30 $\mu$K-arcmin (60 $\mu$K-arcmin) and 40 $\mu$K-arcmin (95 $\mu$K-arcmin), respectively.  The lensing map was checked for systematic effects from the Galaxy, point sources, dust, and instrumental noise bias \citep{2015arXiv150201591P}, which were found to be mostly sub-dominant to the statistical errors.%  More information on the lensing maps can be found in \cite{2015arXiv150201591P}.

\subsection{Galaxy Survey Maps} \label{S:cmass}

We use the CMASS spectroscopic sample from the Sloan Digital Sky Survey (SDSS) III \citep{2011AJ....142...72E} Baryon Oscillations Spectroscopic Survey (BOSS) \citep{2013AJ....145...10D} Data Release 11 (DR11) \citep{2014MNRAS.441...24A,2015ApJS..219...12A}, which was publicly released with the final BOSS data set.  SDSS-III, like SDSS I and II \citep{2000AJ....120.1579Y}, consists of a 2.5 m telescope \citep{2006AJ....131.2332G} with a five-filter ($ugriz$) \citep{1996AJ....111.1748F,2002AJ....123.2121S,2010AJ....139.1628D} imaging camera \citep{1998AJ....116.3040G}, designed to image over one-third of the sky.  Automated pipelines are responsible for astrometric calibration \citep{2003AJ....125.1559P}, photometric reduction \citep{2001ASPC..238..269L}, and photometric calibration \citep{2008ApJ...674.1217P}.  Bright galaxies, luminous red galaxies (LRGs), and quasars are selected for follow-up spectroscopy \citep{2002AJ....124.1810S,2001AJ....122.2267E,2002AJ....123.2945R,2003AJ....125.2276B,2013AJ....146...32S}.  The data used in this survey were acquired between August 1998 and May 2013.

CMASS \citep{2014MNRAS.441...24A} ($z=0.43-0.7$) consists of 690,826 galaxies over an area of 8498 deg$^2$, has a mean redshift of 0.57, and is designed to be stellar-mass-limited at $z>0.45$.  Each spectroscopic sector, or region covered by a unique set of spectroscopic tiles \citep{2011ApJS..193...29A}, was required to have an overall completeness (the fraction of spectroscopic targets that were observed) over 70\% and a redshift completeness (the fraction of observed galaxies with good spectra) over 80\%.  We use these galaxies to construct an overdensity map $\delta_i=(n_i-\bar{n})/\bar{n}$, where $i$ denotes the pixel on the sky.  $n_i=\sum_{j\in {\rm pixel}\,i} w_j$, where $w_j$ is the systematic weight \citep{2014MNRAS.441...24A} of galaxy $j$.  The map is given a HEALPix pixelization with $N_{\rm side}=1024$.  Note that we do not weigh the pixels by their observed area because the HEALPix pixels are much smaller than the observed sectors for which the completeness is computed, and we did not want to introduce extra power due to possible errors in the completeness on small scales.  We also confirm (see Section \ref{S:cmbgalsys}) that including pixel weights have only a small effect on the final result.

\section{Angular Power Spectra} \label{S:cl}
\subsection{Theory}

We model the theoretical galaxy-CMB lensing convergence angular cross-power spectrum and the galaxy clustering angular auto-power spectrum using standard methods.  We assume $\Lambda$CDM with parameters consistent with \emph{Planck} 2013 \citep{2014A&A...571A..16P} and BOSS Data Release 11 \citep{2014MNRAS.441...24A}. We use these models to estimate statistical errors from mocks and systematic corrections to $E_G$ (see Section \ref{S:estim}).  However, our measurement of $E_G$ along with errors from jackknife resampling, which we use in our final result, does not use our power spectrum models and is independent of $\Lambda$CDM. In addition, the corrections we determine from these models are well within statistical error bars.

Using the Limber approximation for small scales ($\ell\gtrsim 10$) and assuming the $\Lambda$CDM model, the galaxy-CMB lensing convergence angular cross-power spectrum can be written as
\begin{eqnarray} \label{E:clkg}
C_\ell^{\kappa g}&=&\frac{3H_0^2\Omega_{m,0}}{2c^2}\int_{z_1}^{z_2}dz\,W(z)f_g(z)\chi^{-2}(z)(1+z)\nonumber\\
&&\times P_{mg}\left[\frac{\ell}{\chi(z)},z\right]\, ,
\end{eqnarray}
where $f_g(z)$ is the galaxy redshift distribution, $W(z)=\chi(1-\chi(z)/\chi_{\rm CMB})$ is the CMB lensing kernel, $\chi(z)\,(\chi_{\rm CMB})$ is the comoving distance out to redshift $z$ (the CMB surface-of-last-scattering redshift $z_{\rm CMB}=1100$), and $P_{mg}(k,z)$ is the matter-galaxy 3D cross-power spectrum as a function of $z$ and wavenumber $k$ \citep{2004PhRvD..70j3501H}.  The cosmological parameters present are the Hubble parameter today $H_0$ and the current matter density parameter $\Omega_{m,0}$.  The galaxy redshift distribution for CMASS is shown in Fig.~1 of \citet{2014MNRAS.441...24A}.  The galaxy clustering angular auto-power spectrum can be written as
\begin{eqnarray} \label{E:clgg}
C_\ell^{gg}=\int_{z_1}^{z_2}dz\,\frac{H(z)}{c}f_g^2(z)\chi^{-2}(z)P_{gg}\left[\frac{\ell}{\chi(z)},z\right]\, ,
\end{eqnarray}
where $H(z)$ is the Hubble parameter at redshift $z$ and $P_{gg}(k,z)$ is the galaxy 3D auto-power spectrum.

\subsection{Mock Galaxy Catalogues from $N$-body Simulations} \label{S:nbody}

We compute the power spectra $P_{mg}(k,z)$ and $P_{gg}(k,z)$ using N-body simulations in order to model both nonlinearities and the occupation of halos with galaxies. The $N$-body simulation runs using the TreePM method \citep{2002JApA...23..185B,White2002,Reid14}. We use 10 realizations of this simulation based on the $\Lambda$CDM model with $\Omega_m= 0.292$ and $h=0.69$.  Although these parameters differ from those from the joint Planck/BOSS analysis, this should not affect the results because $P(k)$ is not so sensitive to the cosmic parameters relative to $C_\ell$. These simulations are in a periodic box of side length 1380$h^{-1}$Mpc and $2048^3$ particles. A friend-of-friend halo catalogue was constructed at an effective redshift of $z=0.55$. This is appropriate for our measurement since the galaxy sample used has effective redshift of 0.57. We use a Halo Occupation Distribution (HOD) \citep{Peacock2000,Seljak2000,Benson2000,White2001,Berlind2002,2002PhR...372....1C} to relate the observed clustering of galaxies with halos measured in the $N$-body simulation. We have used the HOD model proposed in \citet{2014MNRAS.443.1065B} to populate the halo catalogue with galaxies. 
\begin{eqnarray}
\VEV{N_{\rm cen}}_M &= \frac{1}{2} \left[ 1+ \mathrm{erf}\left( \frac{\log M- \log M_{\rm min}}{\sigma_{\log M}}\right) \right] \,\nonumber \\
\VEV{N_{\rm sat}}_M &= \VEV{N_{\rm cen}}_M \left( \frac{M}{M_{\rm sat}} \right)^\alpha \exp \left( \frac{-M_{\rm cut}}{M}\right)
\label{eqn:HOD}\, ,
\end{eqnarray}
where $\VEV{N_{\rm cen}}_M$ is the average number of central galaxies for a given halo mass $M$ and $\VEV{N_{\rm sat}}_M$ is the average number of satellites galaxies. We use the HOD parameter set ($M_{\rm min}=9.319 \times 10^{13} M_\odot/h, M_{\rm sat}=6.729 \times 10^{13} M_\odot/h ,\sigma_{\log M}=0.2, \alpha=1.1, M_{\rm cut}=4.749 \times 10^{13} M_\odot/h$) from \citet{2014MNRAS.443.1065B}. We have populated central galaxies at the center of our halo. The satellite galaxies are populated with radius (distance from central galaxy) distributed as per the NFW profile out to $r_{200}$ and the direction is chosen randomly with a uniform distribution.

\section{Estimators} \label{S:estim}

We estimate $C_\ell^{\kappa g}$ and $C_\ell^{gg}$ along with errors using the \emph{Planck} CMB lensing map and CMASS galaxy map.  We estimate both angular power spectra in 11 flat band-powers that comprise the range $62\leq\ell\leq400$, with each band containing the minimum-variance estimate of the power spectrum over that band. Note that this angular scale range is equivalent to the range 23 Mpc/$h$ $<R_\perp<$ 150 Mpc/$h$, where $R_\perp = 2\pi\chi(z)/\ell$ is the linear scale on the sky corresponding to the angular scale $\ell$ at redshift $z$.  We do not use angular scales $\ell>400$ ($R_\perp<23$ Mpc/$h$) because the CMB lensing convergence at these scales is likely to be contaminated by Gaussian and point-source bias corrections in the lensing estimator \citep{2014A&A...571A..17P}.  We do not use angular scales $\ell<62$ ($R_\perp>150$ Mpc/$h$) because measurements by the BOSS collaboration of $P_{gg}(k)$ at larger scales were shown to be inconsistent between the north and south Galactic caps \citep{2012MNRAS.424..564R}, suggesting the larger-scale measurement could be plagued by systematics.

We estimate $C_\ell^{\kappa g}$ using a pseudo-$C_\ell$ estimator of the form \citep{2011JCAP...03..018L,2014A&A...571A..17P}
\begin{eqnarray}\label{E:pseudocl}
\hat{C}_\ell^{\kappa g}=\frac{1}{(2\ell+1)f_{\rm sky}^{\kappa g}}\sum_{m=-\ell}^\ell g_{\ell m}\kappa_{\ell m}^*\, ,
\end{eqnarray}
where $f_{\rm sky}^{\kappa g}$ is the sky fraction common to the galaxy catalog and the CMB lensing convergence map, $\kappa_{\ell m}$ is the spherical harmonic transform of the CMB lensing convergence field, and $g_{\ell m}$ is the spherical harmonic transform of the galaxy overdensity field.  The error in $\hat{C}_\ell^{\kappa g}$ is estimated as
\begin{eqnarray}\label{E:pseudoclerr}
\sigma^2(\hat{C}_\ell^{\kappa g})=\frac{1}{(2\ell+1)f_{\rm sky}^{\kappa g}}\left[(\hat{C}_\ell^{\kappa g})^2+\hat{D}_\ell^{\kappa\kappa}\hat{D}_\ell^{gg}\right]\, ,
\end{eqnarray}
where $\hat{D}_\ell^{\kappa\kappa}$ and $\hat{D}_\ell^{gg}$ are estimators of the $\kappa$ and galaxy angular auto-power spectra with statistical noise included, given by
\begin{eqnarray}
\hat{D}_\ell^{\kappa\kappa}=\frac{1}{(2\ell+1)f_{\rm sky}^\kappa}\sum_{m=-\ell}^\ell \left|\kappa_{\ell m}\right|^2\, ,
\end{eqnarray}
and
\begin{eqnarray}
\hat{D}_\ell^{gg}=\frac{1}{(2\ell+1)f_{\rm sky}^g}\sum_{m=-\ell}^\ell \left|g_{\ell m}\right|^2\, ,
\end{eqnarray}
where $f_{\rm sky}^\kappa$ and $f_{\rm sky}^g$ are the sky fractions for the CMB lensing convergence map and galaxy catalog, respectively.  We can then use $\hat{C}_\ell^{\kappa g}$ and $\sigma(\hat{C}_\ell^{\kappa g})$ to bin the angular cross-power spectrum.  Note since the lensing field is not Gaussian, least-squares estimates of $C_\ell^{\kappa g}$ will be slightly biased, but not significantly compared to our measurement errors.

We estimate $C_\ell^{gg}$ using a quadratic minimum-variance estimator, a method which has been used in previous estimates \citep{1997PhRvD..55.5895T,2003NewA....8..581P, 2008PhRvD..78d3519H}.  Note we do not estimate $C_\ell^{\kappa g}$ using this method because the required covariance matrix for the CMB lensing convergence is not well-defined.  We estimate $C_\ell^{gg}$ in the same 11 $\ell$-bins used for $C_\ell^{\kappa g}$.     We construct a parameter vector $\mathbf{p}$ that contains all the band-powers for $C_\ell^{gg}$, whose minimum-variance estimator is given by $\hat{\mathbf{p}}=\mathbf{F}^{-1}\mathbf{q}$, where
\begin{eqnarray}
F_{ij}=\frac{1}{2}{\rm tr}\left[\mathbf{C},_i\mathbf{C}^{-1}\mathbf{C},_j\mathbf{C}^{-1}\right]\, ,
\end{eqnarray}
and
\begin{eqnarray}
q_i=\frac{1}{2}\mathbf{x}^T\mathbf{C}^{-1}\mathbf{C},_i\mathbf{C}^{-1}\mathbf{x}\, ,
\end{eqnarray}
are the Fisher matrix and quadratic estimator vector, respectively, $\mathbf{x}$ is the galaxy overdensity map, $\mathbf{C}=\VEV{\mathbf{x}\mathbf{x}^T}$ is the covariance matrix, and $\mathbf{C},_i=\partial\mathbf{C}/\partial p_i$.  Note that $\mathbf{x}$ and $\mathbf{C}$ are given in pixel space.  %The covariance in the power spectrum estimator is given by ${\rm Cov}(p_i,p_j)=F_{ij}$.  
The iterative and stochastic methods used for matrix inversion and trace estimation are described in \citet{2004PhRvD..70j3501H,2007MNRAS.378..852P}.

%We estimate $\beta$ by using the Landy-Szalay estimator \citep{LandySzalay93} to compute a two-dimensional galaxy auto-correlation. We project the galaxy auto-correlation on to the Legendre basis in order to obtain the monopole and quadrupole moments. We fit the monopole and quadrupole moments of the correlation function using Convolution Lagrangian Perturbation Theory (CLPT) and the Gaussian Streaming Model (GSM) \citep{Carlson12,Wang13}. We use scales between 30 $h^{-1}$Mpc to 126 $h^{-1}$Mpc in order to measure $f\sigma_8$ and $b\sigma_8$ \citep{2015MNRAS.453.1754A}, where $f$ is the logarithmic derivative of the growth factor and $b$ is linear galaxy bias.  The RSD parameter is computed by taking the ratio of the measured growth rate and bias $\beta=f/b$.  Although we do not use scales $126<R_\perp<150$ Mpc/$h$ in our $\beta$ measurement, the information in these scales is relatively low due to cosmic variance, and we expect $\beta$ to not be significantly different at these scales.  We do not fit $\beta$ at scales $R_\perp<30$ Mpc/$h$ due to worries about the accuracy of the GSM and the mocks at smaller scales.  Thus, when constructing our measurement of $E_G$, we are able to use our $\beta$ measurement over the full range $23<R_\perp<150$ Mpc/$h$.

The measurement of the redshift space distortions (RSD) parameter $\beta$ is one of the key requirements to measure $E_G$. We estimate $\beta$ by fitting the monopole and quadruple moments of the galaxy auto-correlation function. We use the Landy-Szalay estimator \citep{LandySzalay93} to compute a two-dimensional galaxy auto-correlation. We project the galaxy auto-correlation onto the Legendre basis in order to obtain the monopole and quadrupole moments. We fit the monopole and quadrupole moments of the correlation function using Convolution Lagrangian Perturbation Theory (CLPT) and the Gaussian Streaming Model (GSM) \citep{Carlson12,Wang13}. We measure $f\sigma_8$ and $b\sigma_8$ using scales between 30 $h^{-1}$Mpc to 126 $h^{-1}$Mpc following \cite{2015MNRAS.453.1754A}, where $f$ is the logarithmic derivative of the growth factor and $b$ is linear galaxy bias. We tested our RSD model against various systematics and mocks as described in \cite{2015MNRAS.453.1754A}. We run a Markov Chain Monte Carlo(MCMC) to fit for the galaxy auto-correlation function using COSMOMC \citep{2002PhRvD..66j3511L}. We obtain the likelihood of RSD parameter $\beta$ for each jackknife region by taking the ratio of the measured growth rate and bias $\beta=f/b$. The mean $\beta$ from each jackknife region is then combined to get the final measurement of $\beta$.  Although we do not use scales $126<R_\perp<150$ Mpc/$h$ in our $\beta$ measurement, the information in these scales is relatively low due to cosmic variance, and we expect $\beta$ to not be significantly different at these scales.  We do not fit $\beta$ at scales $R_\perp<30$ Mpc/$h$ as we do not have mocks that can validate the theory model (GSM) at these scales.

\subsection{Error Estimates} \label{S:errest}
We use two methods to determine the errors in $\hat{C}_\ell^{\kappa g}$, $\hat{C}_\ell^{gg}$, and $\hat{\beta}$, namely jackknife resampling and mocks.  Jackknife resampling includes systematics effects naturally; however, the jackknife regions we use, which are all more than 250 Mpc/$h$, may introduce errors in the covariance matrix at the largest scales we sample.  Thus, we also perform a separate error analysis using CMASS mock galaxy catalogs with simulated lensing convergence maps as a check at large scales.

\begin{figure}
\begin{center}
\includegraphics[width=0.5\textwidth]{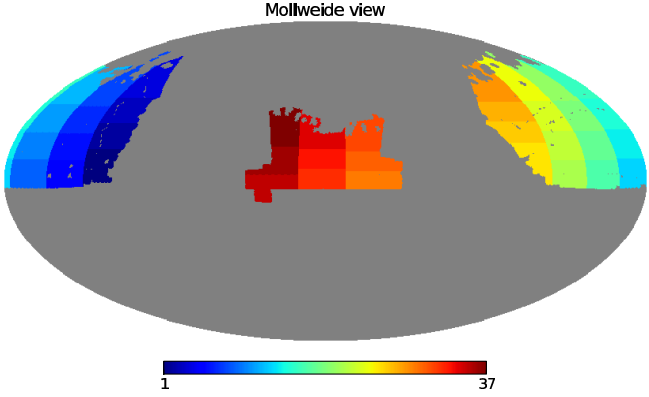}
\caption{\label{F:jack} An equatorial map of the CMASS survey divided into 37 regions used for jackknife resampling.}
\end{center}
\end{figure}

For the first method, we perform jackknife resampling of 37 equal-weight regions of the CMASS survey area, where weight is defined as the effective observed area calculated using the number of random galaxies in CMASS random galaxy maps.  Note that this is not necessarily given by the sky area.  We plot the 37 regions in Fig.~\ref{F:jack}.  Each jackknife region is at least 250 Mpc/$h$ on a side, total weights for regions in the CMASS north galactic cap differ from the CMASS south galactic cap by less than 2\%, and the total weights of each jackknife region differ within a galactic cap by less than 0.8\% (less than 0.1\% for most regions).  We use jackknife resampling to compute expectation values for $\hat{C}_\ell^{\kappa g}$, $\hat{C}_\ell^{gg}$, and $\hat{\beta}$, as well as $\hat{E}_G(\ell)$ and the covariance matrix ${\rm Cov}(E_G)$ for $E_G(\ell)$.  %Note that we only estimate $E_G$ in 6 $\ell$-bins comprising the range $100\leq\ell\leq250$ due to concerns about large-scale systematics and nonlinear clustering.  %The figures for the angular power spectra and $E_G$ are shown in the main text.  We show the correlation matrix for $E_G$ using jackknife resampling in figure (\ref{F:covjack}), where we see that the first three $\ell$-bins ($100\leq\ell\leq160$) are correlated with each other and with the 6th bin ($200\leq\ell\leq250$).%  The correlations could be due to foregrounds or random fluctuations in the ; more study would be needed to confirm this.

The second method computes $\hat{C}_\ell^{\kappa g}$, $\hat{C}_\ell^{gg}$, $\hat{\beta}$, and $\hat{E}_G(\ell)$ using the full \emph{Planck} and CMASS surveys, and the covariances for all four quantities are determined using simulations and mocks.  For the CMB lensing field, we simulate 100 convergence maps, in which each map is a Gaussian field with the correct signal and noise power spectra and mask provided by \emph{Planck}.  For the galaxies, we use 100 CMASS mock catalogs \citep{2013MNRAS.428.1036M}.  The halo occupation density used to construct these mocks, presented in \citet{2013MNRAS.428.1036M}, was significantly updated recently \citep{2014MNRAS.443.1065B}. This is reflected in that $C_\ell^{gg}$ for the mocks from \citet{2013MNRAS.428.1036M} are significantly lower than the data on small scales, the data which was fitted to determine the HOD in \citet{2014MNRAS.443.1065B}.  We remedy this by rescaling $C_\ell^{gg}$ for all the mocks equally such that the average $C_\ell^{gg}$ of the mocks matches $C_\ell^{gg}$ from the data.  Also, the lensing simulations were not constructed to be correlated with the mocks, but we do not expect this to be important because the CMB lensing-galaxy correlation should only contribute 1-2\% of the errors in $C_\ell^{\kappa g}$. Also, since the galaxy mocks are not correlated with the $\kappa$ simulations, there is no need to rescale $C_\ell^{\kappa g}$.  %We show the correlation matrix for $E_G$ using simulations and mocks in figure (\ref{F:covmock}), where we see that $\ell$-bin pairs 1-4, 2-5, 2-6, and 3-6 are correlated.%, possibly due to the wide CMASS redshift distribution.

Assuming the GR case where $E_G$ is independent of $\ell$, we construct a likelihood function given by
\begin{widetext}
\begin{eqnarray}
\mathcal{L}(E_G)&\propto&\exp\left\{-\frac{1}{2}\sum_{\ell,\ell'}[\hat{E}_G(\ell)-E_G][{\rm Cov}(E_G)]_{\ell,\ell'}^{-1}[\hat{E}_G(\ell')-E_G]\right\}\, ,
\end{eqnarray}
\end{widetext}
from which we determine the maximum likelihood value for $E_G$ along with statistical errors.  In order to correct the bias to $[{\rm Cov}]^{-1}$ due to using a finite number of jackknives/mocks, we multiply $[{\rm Cov}]^{-1}$ by $(1-D)$, in which
\begin{eqnarray}
D=\frac{n_b+1}{n_s-1}\, ,
\end{eqnarray}
where $n_b$ is the number of bins for which we estimate the covariance matrix, and $n_s$ is the number of samples \citep{2013arXiv1312.5490P,2014IAUS..306...99J}.  Thus, $D=7/36$ for the jackknives and $D=7/99$ for the simulations/mocks, although we acknowledge that the scaling for the jackknives could be inaccurate at larger scales due to the size of the jackknife regions.  However, this does not appear to make the jackknife results much different from that of the simulations/mocks.%  We plot the likelihood for both methods in figure (\ref{F:like}), where we see that the likelihood from the jackknives is broader than that from the simulations/mocks.  This result leads us to believe that we are likely overestimating the errors due to jackknives, and thus the jackknife errors are the more conservative choice.

\subsection{Systematic Corrections to $E_G$} \label{S:syscor}

Our estimator for $E_G$ in Eq.~\ref{E:egth} is not unbiased due to scale-dependent clustering bias as well as a mismatch between the CMB lensing kernel and the redshift distribution of CMASS galaxies.  We apply systematic corrections to our $E_G$ estimator to debias our result, which we outline in this subsection.  These correction factors are similar in purpose to those applied to the first $E_G$ estimate in \citet{2010Natur.464..256R}, although their kernel and effective redshift corrections are combined in our the kernel mismatch correction.

We derive $\Gamma$ in Eq.~\ref{E:egth} by relating $C_\ell^{\kappa g}$ and $C_\ell^{gg}$ in Eqs.~\ref{E:clkg} and \ref{E:clgg} and then setting $\Gamma$ such that the expectation value of the resulting expression for $E_G$ is consistent with Eq.~\ref{E:egthk}.  It can be shown that by removing the appropriate functions from the integrands which are slowly varying compared to $f_g^2(z)$, the correct expression for $\Gamma$ is
\begin{eqnarray}\label{E:gamma}
\Gamma=\frac{2c}{3H_0}\left[\frac{E(z)f_g(z)}{W(z)(1+z)}\right]\, ,
\end{eqnarray}
where $E(z)=H(z)/H_0$. The approximations required to produce this expression are not perfect, causing $E_G$ measured using Eq.~\ref{E:egth} to slightly deviate from the true value of $E_G$.  We correct this systematic effect by multiplying $\Gamma$ by $C_\Gamma$, given by
\begin{eqnarray}
C_\Gamma(\ell)=\frac{W(z)(1+z)}{2f_g(z)}\left[\frac{c}{H(z)}\right]\frac{C_\ell^{mg}}{Q_\ell^{mg}}\, ,
\end{eqnarray}
where $Q_\ell^{mg}$ and $C_\ell^{mg}$ are defined as
\begin{eqnarray}
Q_\ell^{mg}&\equiv&\frac{1}{2}\int_{z_1}^{z_2}dz\,W(z)f_g(z)\chi^{-2}(z)(1+z)\nonumber\\
&&\times P_{mg}\left[\frac{\ell}{\chi(z)},z\right]\, ,
\end{eqnarray}
and
\begin{eqnarray}
C_\ell^{mg}\equiv \int_{z_1}^{z_2}dz\,\frac{H(z)}{c}f_g^2(z)\chi^{-2}(z)P_{mg}\left[\frac{\ell}{\chi(z)},z\right]\, .
\end{eqnarray}
%and $f(z)\simeq[\Omega_m(z)]^{0.55}$ is the growth rate.

Another systematic correction concerns the clustering bias.  Specifically, while $\beta$ is computed using the linear bias, the angular power spectra are computed over a range of scales, including small, non-linear scales where the clustering bias is scale-dependent.  This causes the clustering bias factors in $E_G$ to not fully cancel.  This systematic effect is corrected by multiplying $E_G$ by $C_b$, where
\begin{eqnarray}
C_b(\ell)=\frac{C_\ell^{gg}}{b_{\rm lin}C_\ell^{mg}}\, .
\end{eqnarray}

\begin{figure}
\begin{center}
\includegraphics[width=0.5\textwidth]{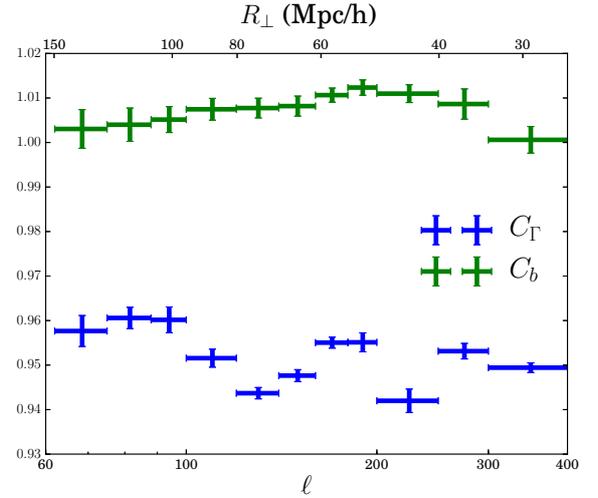}
\caption{\label{F:egsys}  Correction factors applied to $E_G$ due to $\Gamma$ (solid) and bias (dashed).  These correction factors were determined from N-body simulations.}
\end{center}
\end{figure}

We plot these corrections to $E_G$ in Fig.~\ref{F:egsys} for the same 11 $\ell$-bins used to compute $E_G$ in section \ref{S:estim}.  We see that the $\Gamma$ correction is approximately 6\% from unity with $\pm1$\% variation, while the bias correction is only 1\% from unity with little variation.  The errors are due to the fluctuations in the 10 N-body simulations used to calculate the power spectra.  The size of $C_\Gamma$ is due to the kernels of $Q_\ell^{mg}$ and $C_\ell^{mg}$ peaking at different redshifts, and the wiggles are due to baryonic acoustic oscillations.   Note that we did not include uncertainties in cosmological parameters into the errors.  By combining the errors for these corrections over the scale range, we estimate a systematic error of 1.2\%.

\begin{figure}
\begin{center}
\includegraphics[width=0.5\textwidth]{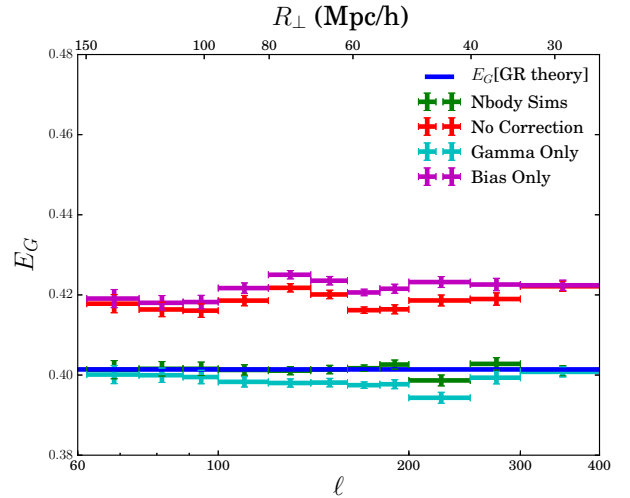}
\caption{\label{F:egsystest}  Test of $E_G$ correction factors $C_\Gamma$ and $C_b$ on N-body simulations (green crosses) compared to the fiducial value of $E_G$ (solid).  We also show $E_G$ without corrections (red crosses) and with only the $\Gamma$ correction (cyan crosses) or the scale-dependent bias correction (magenta crosses).}
\end{center}
\end{figure}

We test our corrections by computing $E_G(\ell)$, both with and without corrections, based on the N-body simulations and comparing them with the fiducial value.  In Fig.~\ref{F:egsystest}, we see that the result matches well with the fiducial value.  It is possible that the modeling of the clustering in the N-body simulations and the HOD could affect the corrections, particularly $C_b$.  Incorrect modeling of the redshift distribution could also affect the corrections, particularly $C_\Gamma$.  The statistical error on our final $E_G$ estimate is large enough such that this should not be an issue, but this could affect upcoming $E_G$ measurements that are more precise, requiring more precise modeling of the corrections using simulations.

%As with the statistical errors, we also construct a likelihood function for these systematic errors, given by
%\begin{widetext}
%\begin{eqnarray}
%\mathcal{L}(C_\Gamma C_b)\propto\exp\left\{-\frac{1}{2}\sum_{\ell,\ell'}[\hat{C}_\Gamma(\ell) \hat{C}_b(\ell)-C_\Gamma C_b][{\rm Cov}(C_\Gamma C_b)]_{\ell,\ell'}^{-1}[\hat{C}_\Gamma(\ell') \hat{C}_b(\ell')-C_\Gamma C_b]\right\}\, ,
%\end{eqnarray}
%\end{widetext}
%where ${\rm Cov}(C_\Gamma C_b)$ is the covariance matrix for $C_\Gamma C_b$ determined from simulations.  We also rescale ${\rm Cov}^{-1}$ by $(1-D)$, where $D=7/9$ [see section \ref{S:errest}].  From our likelihood function, we find $C_\Gamma C_b=0.947\pm0.00595$, implying an error of 0.6\%.

\section{Results} \label{S:results}

\begin{figure}[t!]
\begin{center}
\includegraphics[width=0.5\textwidth]{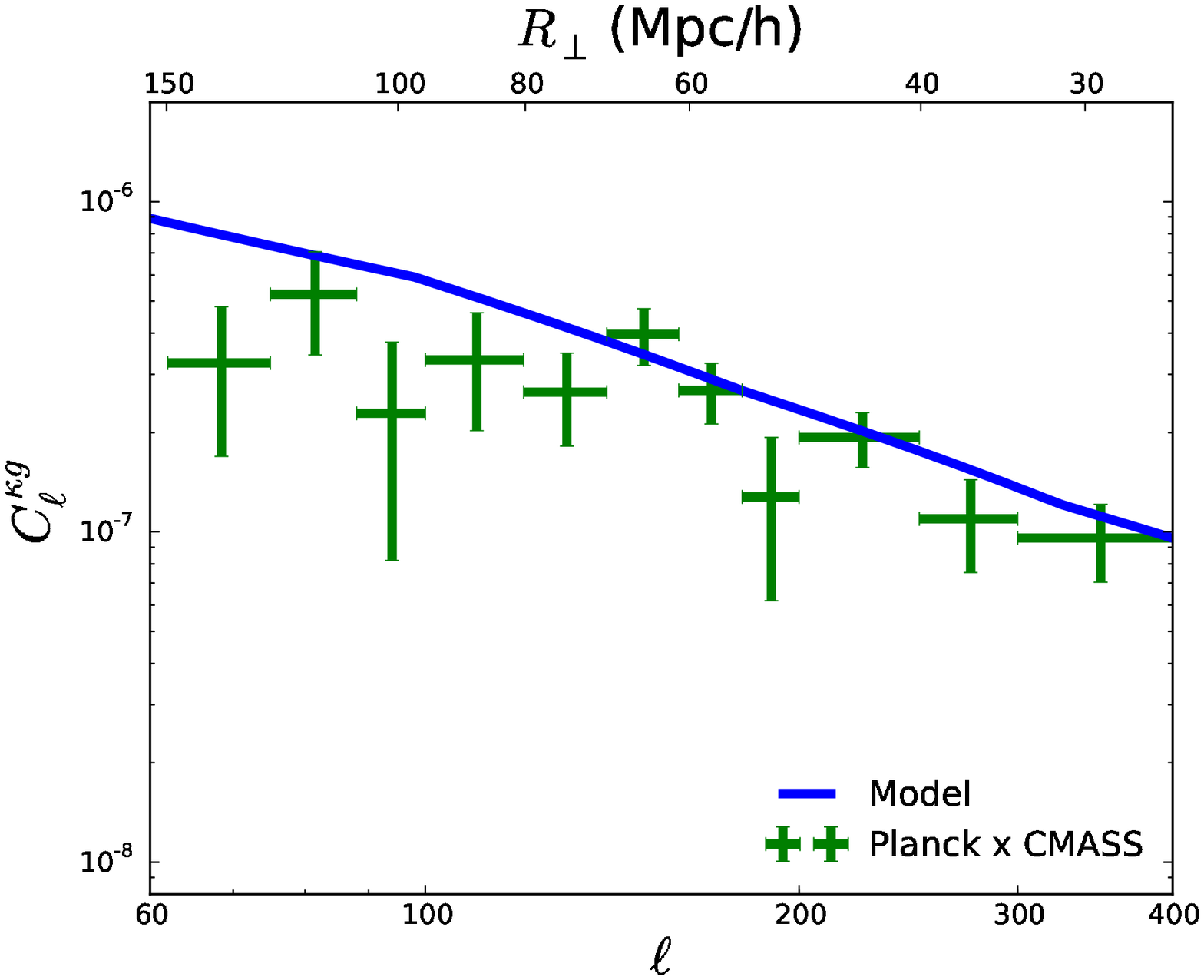}
\includegraphics[width=0.5\textwidth]{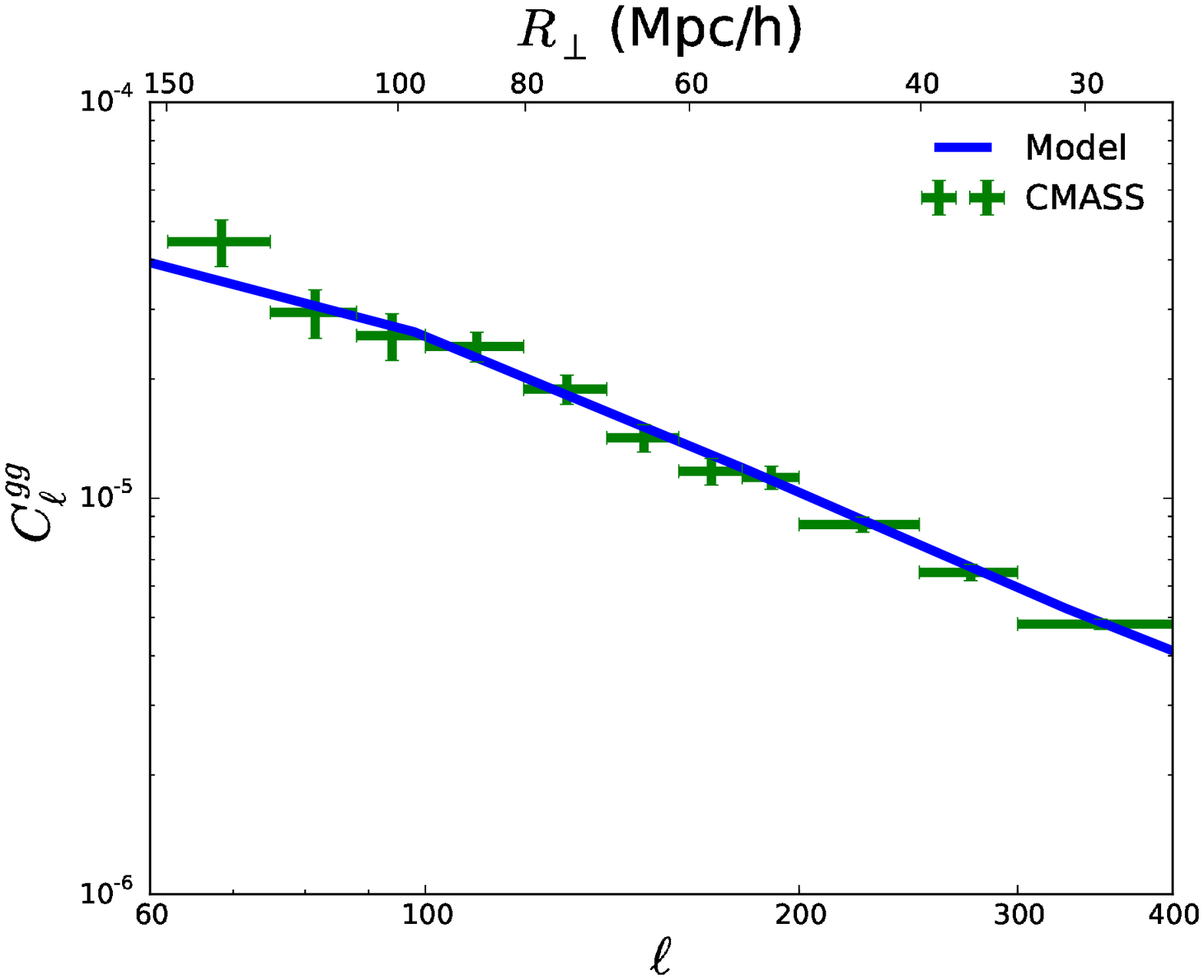}
\caption{\label{F:cl} Observed angular power spectra (crosses) for galaxy-CMB lensing (\emph{top}) and galaxy clustering (\emph{bottom}) with 1$\sigma$ errors using the CMASS galaxy sample and the \emph{Planck} CMB lensing map.  In both panels, we show $\ell$ on the lower horizontal axis and $R_\perp$, the corresponding linear scale projected onto the sky, on the upper horizontal axis.  The errors were derived using jackknife resampling of 37 equally weighted regions in the CMASS survey.  Our galaxy angular power spectrum measurement is consistent with theoretical models (solid lines) derived from N-body simulations, while our galaxy-CMB lensing angular cross-power spectrum is low yet consistent with other measurements, \emph{e.g.}~\citet{2015A&A...584A..53K}.  We discuss possible causes for this deficit in Sec.~\ref{S:results}.}
\end{center}
\end{figure}

\begin{figure}
\begin{center}
\includegraphics[width=0.5\textwidth]{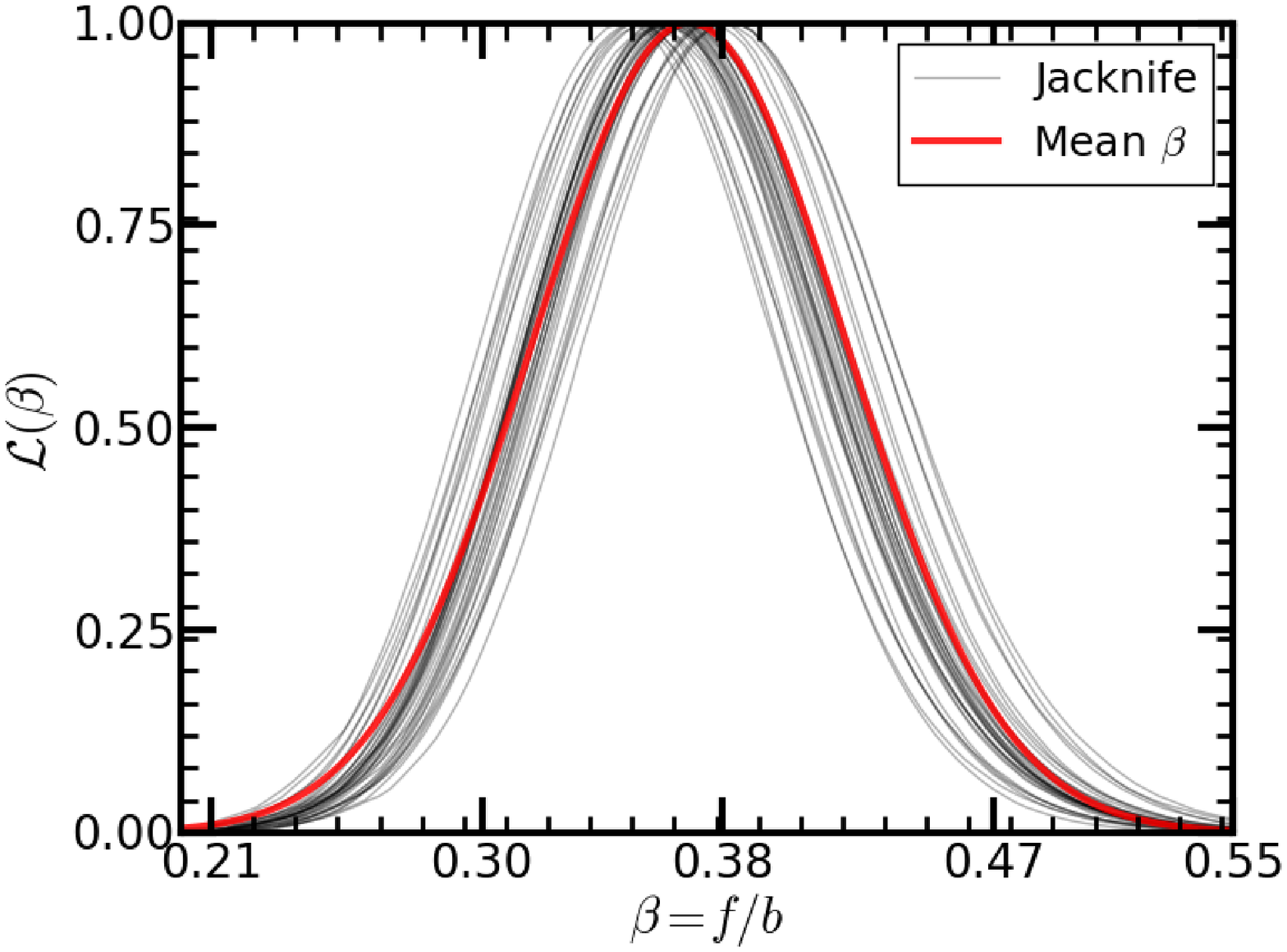}
\includegraphics[width=0.5\textwidth]{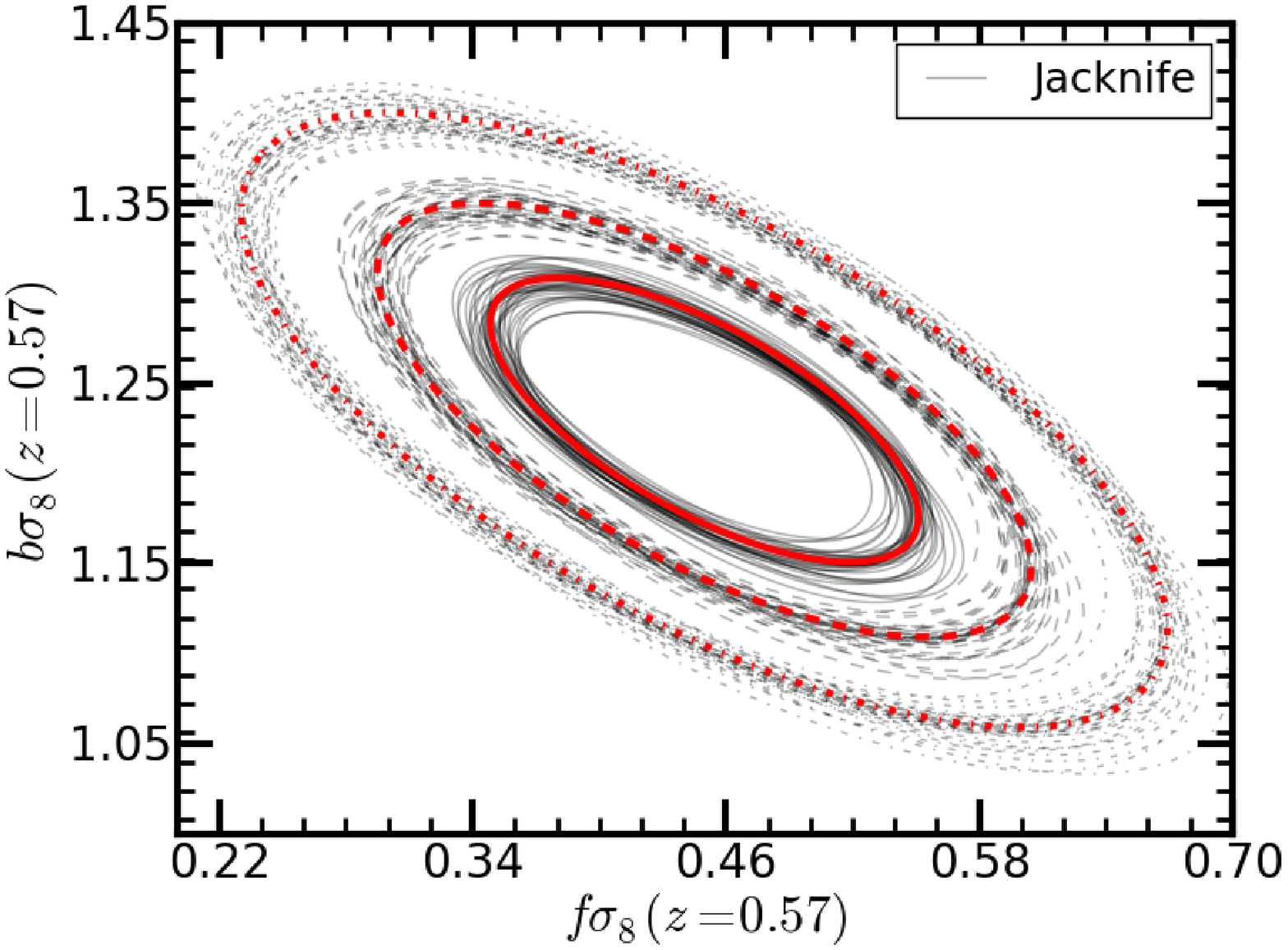}
\caption{\label{fig:betalike} The top plot shows the likelihood of $\beta$ and the bottom plot shows our constraint in the $b\sigma_8-f\sigma_8$ plane. The black lines are the likelihood obtained for individual jackknife regions and the red lines are our likelihood obtained by combining the mean of individual jackknife region. These plots also shows that our jackknife sampling is unbiased in estimating the parameters.}
\end{center}
\end{figure}

We show in Fig.~\ref{F:cl} the angular power spectra for galaxy-CMB lensing, $C_\ell^{\kappa g}$, and galaxy clustering, $C_\ell^{gg}$, which we estimate from the \emph{Planck} CMB lensing map and the CMASS galaxy number density maps using jackknife resampling.  It is evident that the measured $C_\ell^{gg}$ is consistent with the theoretical prediction from $\Lambda$CDM combined with the HOD model.  However, the measured $C_\ell^{\kappa g}$ is a bit lower at large scales than the theoretical prediction.  Specifically, we find a cross-correlation amplitude of $A=0.754\pm0.097$, which is low but consistent with the value reported in \citet{2015A&A...584A..53K}, $A=0.85^{+0.15}_{-0.16}$, for \emph{Planck} cross-correlated with the CFHTLens\footnote{http://cfhtlens.org} galaxy sample.  Note that this low value of $A$ is also inconsistent with values of $A>1$ favored by the \emph{Planck} CMB temperature and polarization maps alone \citep{2015arXiv150201589P,2015arXiv150702704P}.  We also perform jackknife resampling for the RSD parameter, finding $\beta=0.368\pm0.046$.  The full results for $\beta$, including the likelihood and the measurements of $b\sigma_8$ and $f\sigma_8$, are shown in Fig.~\ref{fig:betalike}.

%In any case, we do not use power spectrum measurements on angular scales $\ell<100$ in our final results, so this relative lack of power should not impact our conclusions.
% The errors for each power spectrum were computed by jackknife resampling 37 equally weighted regions that comprise the CMASS galaxy survey.

%\textbf{Large-scale structure measurements from the CMB and $\sim$700,000 galaxies.}

\begin{figure}
\begin{center}
\includegraphics[width=0.5\textwidth]{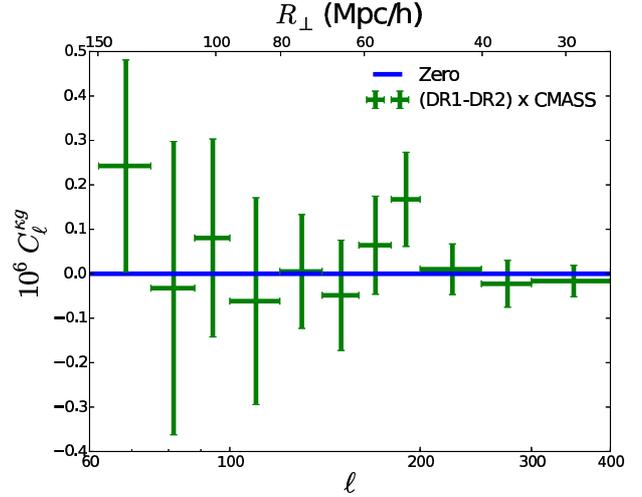}
\caption{\label{F:clkgdiff} Observed angular cross-power spectrum (crosses) with 1$\sigma$ errors between the CMASS galaxy sample and the difference map between the \emph{Planck} 2013 (DR1) and 2015 (DR2) CMB lensing maps.  We show $\ell$ on the lower horizontal axis and $R_\perp$, the corresponding linear scale projected onto the sky, on the upper horizontal axis.  The angular cross-power spectrum measurements is consistent with a null result (solid line).}
\end{center}
\end{figure}

\begin{figure}
\begin{center}
\includegraphics[width=0.5\textwidth]{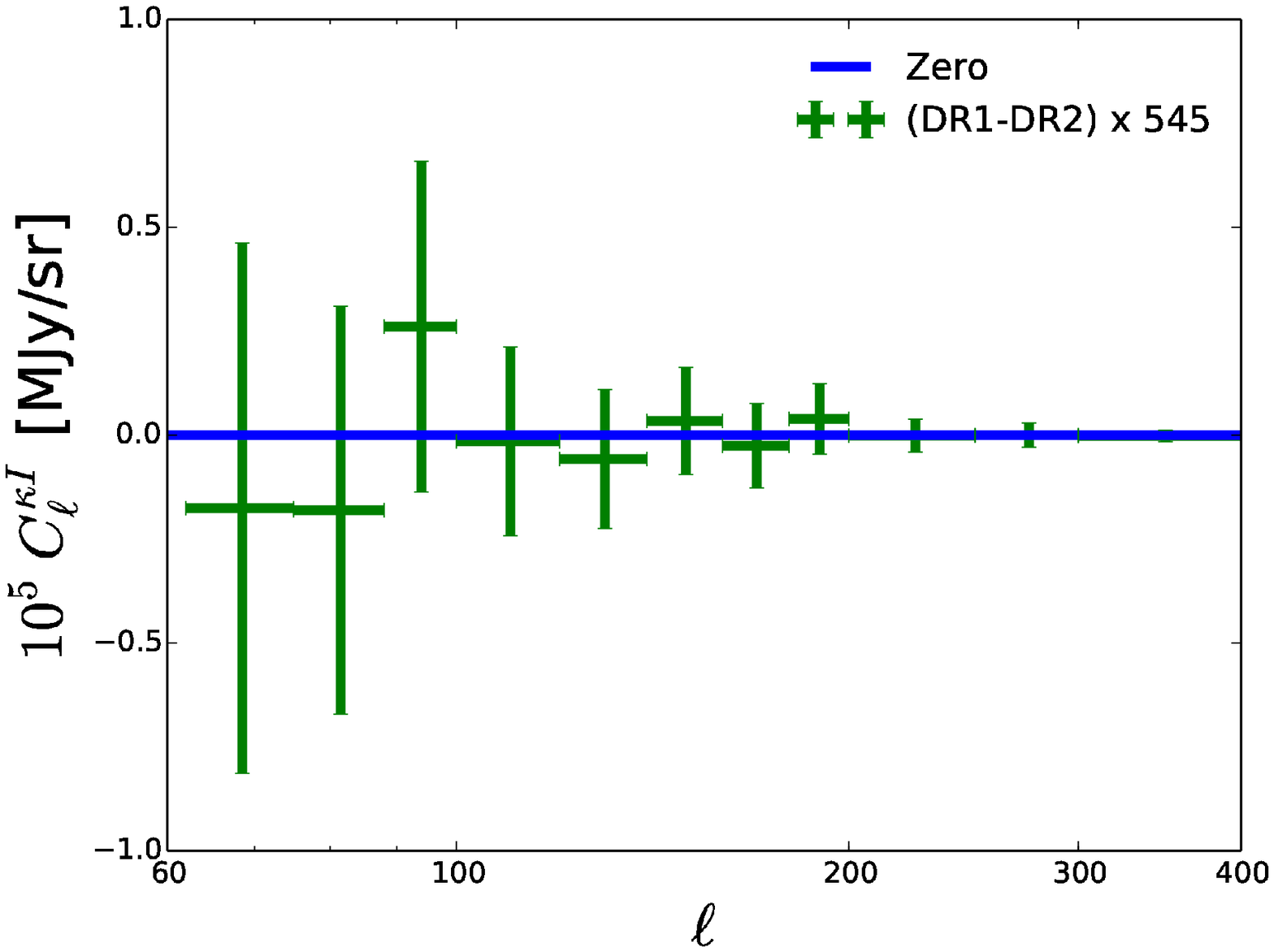}
\caption{\label{F:clk545diff} Observed angular cross-power spectrum (crosses) with 1$\sigma$ errors between the \emph{Planck} 545 GHz map (dust-dominated) and the difference map between the \emph{Planck} 2013 (DR1) and 2015 (DR2) CMB lensing maps.  The format is similar to Fig.~\ref{F:clkgdiff}.  The angular cross-power spectrum measurements is consistent with a null result (solid line).}
\end{center}
\end{figure}

\begin{figure}
\begin{center}
\includegraphics[width=0.5\textwidth]{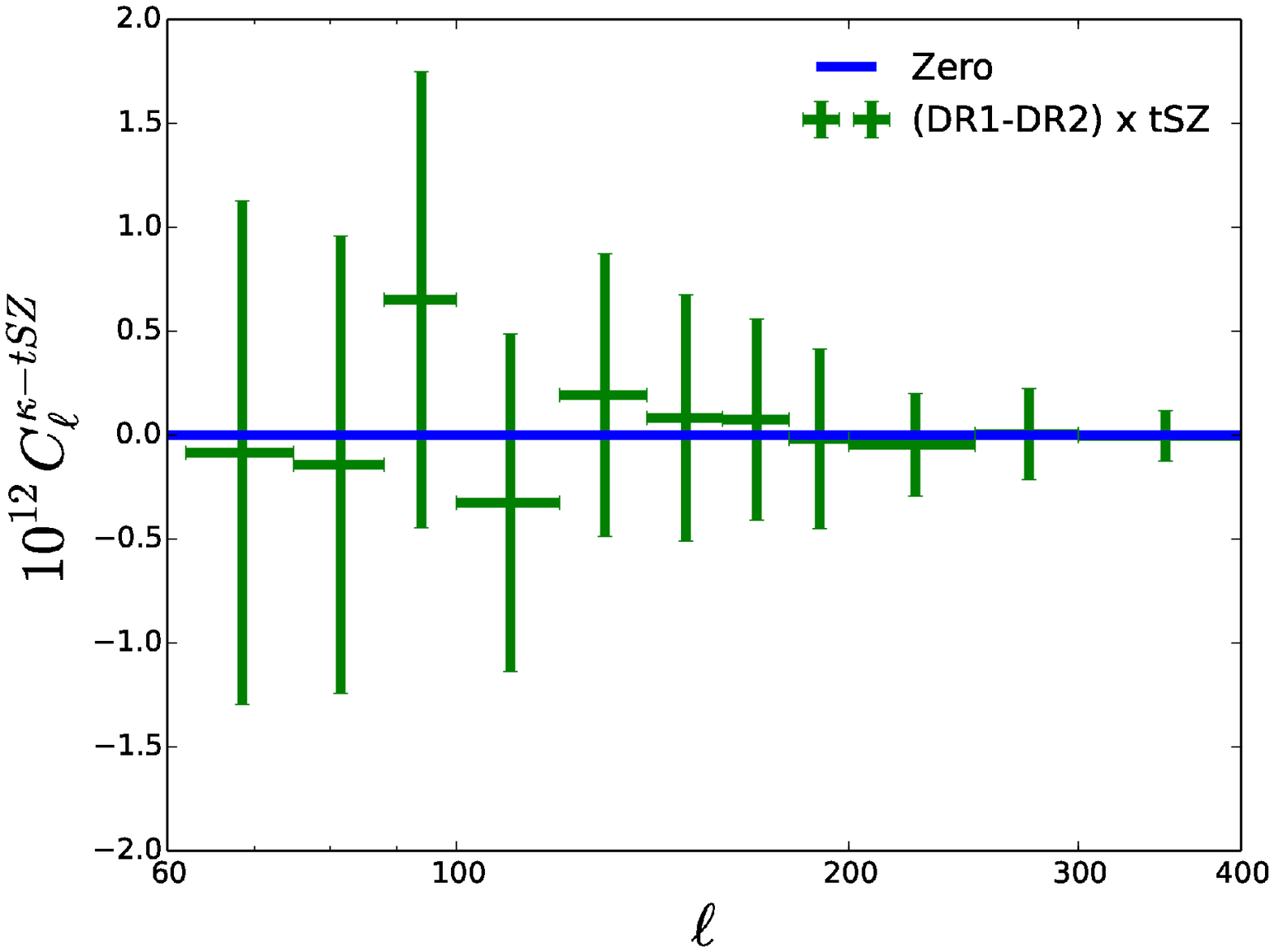}
\caption{\label{F:clkszdiff} Observed angular cross-power spectrum (crosses) with 1$\sigma$ errors between the \emph{Planck} SZ $y$ map and the difference map between the \emph{Planck} 2013 (DR1) and 2015 (DR2) CMB lensing maps.  The format is similar to Fig.~\ref{F:clkgdiff}.  The angular cross-power spectrum measurements is consistent with a null result (solid line).}
\end{center}
\end{figure}

We considered whether the deficit in $C_\ell^{\kappa g}$ at large scales could be due to a systematic effect introduced in the latest lensing map.  Recent work has suggested there may be tension between the \emph{Planck} CMB lensing maps from 2013 and 2015 \citep{2015arXiv150203405O,2015PhRvD..92f3517L,2015A&A...584A..53K}.  In particular, the galaxy cross-correlation with the \emph{Planck} 2015 CMB lensing map appears to measure a smaller clustering bias than the 2013 map, suggesting that the 2013 CMB lensing map may have produced a cross-correlation more consistent with our $C_\ell^{\kappa g}$ model on these scales.  We test this by taking the difference map between the \emph{Planck} 2015 and 2013 CMB lensing maps and cross-correlating with the CMASS map, the \emph{Planck} 545 GHz map (dust-dominated), and the Sunyaev-Zeldovich (SZ) Compton-$y$ map \citep{2015arXiv150201596P}.  In all three cases (see Figs.~\ref{F:clkgdiff}-\ref{F:clkszdiff}) we find the cross-correlations are consistent with zero, suggesting that the \emph{Planck} 2015 and 2013 CMB lensing maps are equivalent, and that any contamination must be common to both maps. It is possible that $C_\ell^{\kappa g}$ could be correlated with the scanning direction, and that lensing convergence maps for separate surveys with different scanning strategies could reveal a discrepancy.  Testing this would require constructing lensing convergence maps for partial surveys, which we leave for future work.

Previous work has also shown \citep{2016MNRAS.456.3213G} that the large-scale $C_\ell^{\kappa g}$ deficit is also present in the cross-correlation between the Dark Energy Survey \citep{2005astro.ph.10346T} Science Verification galaxy sample and the South Pole Telescope CMB lensing map \citep{2015ApJ...810...50S}, which suggests the source of this deficit is not unique to the \emph{Planck} CMB maps.  Thus, it appears that the source of this deficit may very well be astrophysical or cosmological.  The deficit could be caused by thermal SZ contamination, in that the SZ increases the variance in the CMB map, which the lensing estimator interprets as an ``anti-lens.''  Unfortunately, thermal SZ was not removed from the \emph{Planck} SMICA maps \citep{2015arXiv150205956P}.  Recent work \citep{2014ApJ...786...13V} showed that the CMB lensing-galaxy cross-correlation could be biased due to contamination from thermal SZ and the cosmic infrared background (CIB), though the predicted magnitudes of the biases ($\sim 4-6$\%) are too small to explain the deficit.  Also, the lack of evidence for contamination could be due to a lack of power spectrum sensitivity instead of a lack of contamination.  Of course, a combination of causes could also explain the discrepancy.  In addition, other analyses have claimed an excess ($A>1$) galaxy-CMB lensing correlation \citep{2015ApJ...802...64B,2015arXiv151105116B} in contradiction to the deficit seen in the previously mentioned claims.  More research is needed to determine the nature of this deficit; however, we consider this beyond the scope of our investigation and leave this for future work.

%One possible astrophysical explanation for the deficit is a high $\sigma_8$ measurement.  The clustering bias determined from CMASS data \citep{2013MNRAS.428.1036M,2014MNRAS.443.1065B} was set such that the right combination $b\sigma_8$ would produce the measured power spectrum; however, the lensing-galaxy cross-correlation scales as $b\sigma_8^2$, such that a $\sigma_8$ 30\% lower could match our $C_\ell^{\kappa g}$ and $C_\ell^{gg}$.  Note that a lower $\sigma_8$ should not affect our measurement of $E_G$ since $\sigma_8$ cancels in the expression.

\begin{figure}
\begin{center}
\includegraphics[width=0.5\textwidth]{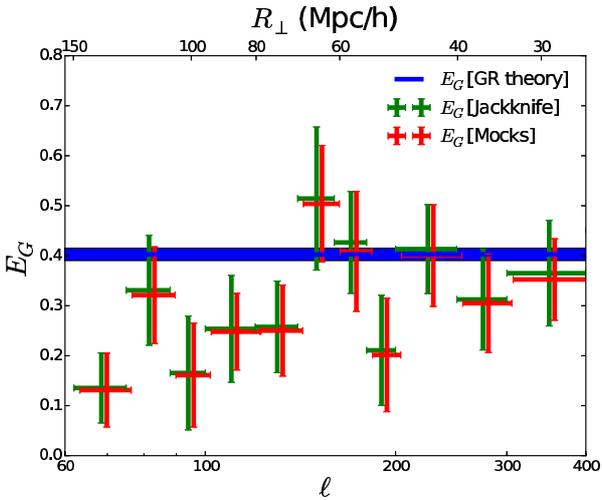}
\caption{\label{F:egv} $E_G$ measurements with 1$\sigma$ errors using the CMASS galaxy sample and the \emph{Planck} CMB lensing map.  The horizontal axes are described in the caption for Fig.~\ref{F:cl}.  We show estimates using jackknife resampling of the full sample [green crosses; see Fig.~\ref{F:cl}] and estimates using the full sample with errors computed from 100 CMASS mock galaxy catalogues and Gaussian simulations of the lensing convergence field (red crosses).  The blue region shows the GR prediction $E_G=0.402\pm0.012$ with the error determined from the likelihood from \emph{Planck} and BOSS measurements.  Averaging the $E_G$ values from jackknife resampling over all scales, we find $E_G=0.243\pm0.060$ ($1\sigma$), a 2.6$\sigma$ deviation from GR.}  %The turquoise line shows the $f(R)$ gravity prediction for the upper limit of the parameter $B_0$ \citep{2007PhRvD..75d4004S}, proportional to the $f(R)$ curvature today.}%  The shaded purple region shows the $E_G$-space of chameleon gravity with parameters\cite{2008PhRvD..78b4015B} $B_0=0.4$, $s=4$, and the range $\beta_1=0.8-1.6$.  The shaded orange region shows the same but for $\beta_1=0.7-0.8$, which is disfavored by our measurement at over 99\% confidence.  Note that an MCMC analysis is necessary to produce a constraint for $\beta_1$ marginalized over $B_0$ and $s$.}
\end{center}
\end{figure}

The power spectra, $C_\ell^{\kappa g}$ and $C_\ell^{gg}$, and $\beta$ are combined using Eq.~\ref{E:egth} to compute $E_G(\ell)$ within 11 $\ell$-bins comprising the angular modes $\ell=62-400$ ($23<R_\perp<150$ Mpc/$h$), which we present in Fig.~\ref{F:egv}.  Note that we probe scales much larger than the previous measurements using galaxy lensing \citep{2010Natur.464..256R,2016MNRAS.456.2806B}.  The range in $\ell$ was chosen to avoid observational systematic effects on large scales \citep{2012ApJ...761...14H,2012MNRAS.424..564R,2014MNRAS.437.1109R} and lensing noise bias on small scales \citep{2014A&A...571A..17P}.  The low values of $E_G$ are attributable to the deficit in $C_\ell^{\kappa g}$, while $E_G$ in the first $\ell$-bin is even lower due to its excess $C_\ell^{gg}$.  The covariance matrix for $E_G(\ell)$ over the 11 $\ell$-bins was computed using jackknife resampling.  Taking the average of $E_G(\ell)$ over $\ell$, while accounting for the covariance matrix, we find $E_G=0.243\pm0.060$ ($1\sigma$). This is a measurement with 25\% statistical errors, over two times larger than forecasts \citep{2015MNRAS.449.4326P} mainly due to the low expectation value we find relative to GR and correlations between $E_G$ estimates at different angular scales, possibly due to systematic foregrounds.   Repeating the $E_G$ estimation using the full CMB lensing and galaxy maps with an $E_G$ covariance matrix produced from the CMASS mock galaxy catalogues \citep{2013MNRAS.428.1036M} and Gaussian simulations of the lensing convergence field gives us a similar result $E_G=0.269\pm0.047$, which is consistent with the result from jackknife resampling.  Since the results using jackknife resampling have larger errors than those using mocks, we choose the more conservative jackknife results as our main GR constraint.

The general relativistic prediction for $E_G$ is given by $\Omega_{m,0}/f(z)=0.402\pm0.012$ at redshift $z=0.57$, based on estimates of the cosmological parameters \citep{2014A&A...571A..16P} by the \emph{Planck} satellite and the BOSS measurements of the baryon acoustic oscillations (BAO) scale.  There is tension between the value from general relativity and our measurement, on the order of 2.6$\sigma$.  We test GR at scales three times as large as those probed in the previous $E_G$ measurements using galaxy-galaxy lensing \citep{2010Natur.464..256R,2016MNRAS.456.2806B}, and it is at these larger scales that this deviation appears.  Specifically, our averaged $E_G$ measurement deviates from the GR value by more than 1$\sigma$ when scales greater than 80 Mpc/$h$ ($\ell<150$) are included.  However, there are still unanswered questions regarding the nature of the deficit in $C_\ell^{\kappa g}$.  Thus, we do not claim significant evidence for a departure from general relativity.

In \citet{2015MNRAS.449.4326P}, the authors derive $E_G$ for $f(R)$ gravity \citep{2004PhRvD..70d3528C}, which is given by
\begin{eqnarray}
E_G^{\rm fR}(k,z)=\frac{1}{1-B_0a^{s-1}/6}\frac{\Omega_{m,0}}{f^{\rm fR}(k,z)}\, ,
\end{eqnarray}
where $B_0$ \citep{2007PhRvD..75d4004S,2008PhRvD..78b4015B} is a free parameter which is related to the Compton wavelength of an extra scalar degree of freedom and is proportional to the curvature of $f(R)$ today, $s$=4 for models that follow $\Lambda$CDM, and $f^{\rm fR}(k,z)$ is the $f(R)$ growth rate, which is scale-dependent \citep{2011JCAP...08..005H}.  Current measurements limit $B_0<1.36\times10^{-5}$ (1$\sigma$) \citep{2015PhRvD..91f3008X,2015PhRvD..91j3503B,2016MNRAS.456.3743A}. $f(R)$ gravity would be indistinguishable by eye from the GR curve in Fig.~\ref{F:egv}, suggesting that we cannot constrain it further using our measurement. The relative constraining power of the RSD measurement alone \citep{2016MNRAS.456.3743A} compared to our measurement is partially due to the use of 6 LSS surveys in the growth rate analysis as compared to our use of one survey in our $E_G$ analysis.  In addition, most of our constraining power is degraded because of the relatively low signal-to-noise ratio of the lensing measurement.  Future CMB surveys such as Advanced ACTPol \citep{2016JLTP..tmp..144H} or possibly the Primordial Inflation Explorer (PIXIE) \citep{2011JCAP...07..025K} with high sensitivities and angular resolution combined with upcoming large-area galaxy surveys with high number densities and moderate redshift precision, along with better control of systematics, should be much more competitive with growth rate measurements without the clustering-bias degeneracy that the growth rate exhibits \citep{2015MNRAS.449.4326P}.  These upcoming $E_G$ measurements should also be capable of differentiating between GR and other gravity models.

\section{CMB Lensing and Galaxy Systematics} \label{S:cmbgalsys}

We consider contamination from dust emission and point sources, which could correlate with both the CMB lensing map and our galaxy sample, thus biasing $C_\ell^{\kappa g}$.  Specifically, for both the CMB lensing map and our galaxy sample we construct 6 cross-correlations, one with a dust map and 5 with point-source maps from \emph{Planck}, using a pseudo-$C_\ell$ estimator similar to Eqs.~\ref{E:pseudocl} and \ref{E:pseudoclerr}.  To trace dust emission, we use the \citet{1998ApJ...500..525S} galactic extinction map using infrared emission data from the Infrared Astronomy Satellite (IRAS) and the Diffuse Infrared Background Experiment (DIRBE).  Three point-source overdensity maps are constructed from the \emph{Planck} Catalog of Compact Sources \citep{2014A&A...571A..28P} at frequencies 100, 143, and 217 GHz.  We also consider the \emph{Planck} SZ Catalog \citep{2015arXiv150201598P} of sources detected through the SZ effect \citep{1980ARA&A..18..537S}, as well as the \emph{Planck} Catalog of Galactic Cold Clumps \citep{2015arXiv150201599P}.%  We plot the cross-correlations in Fig.~\ref{F:cmbsys} with the expectation values errors constructed using a pseudo-$C_\ell$ estimator similar to Eqs.~\ref{E:pseudocl} and \ref{E:pseudoclerr}, finding that the main systematics that could significantly bias $C_\ell^{\kappa g}$ are the compact sources and the SZ sources.

\begin{figure}
\begin{center}
\includegraphics[width=0.5\textwidth]{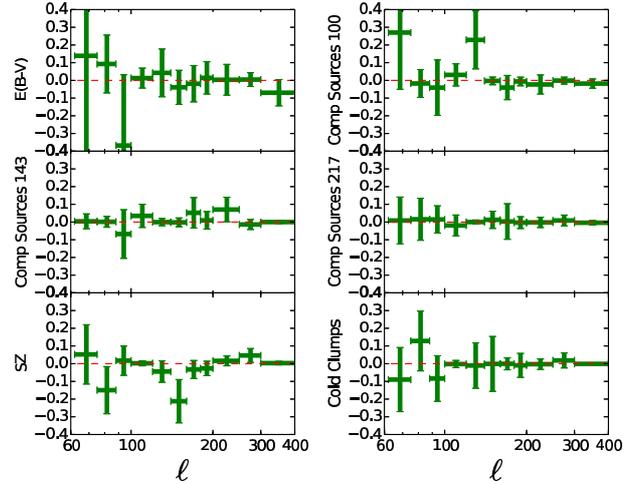}
\caption{\label{F:cmbsys}  The estimated bias to $C_\ell^{\kappa g}$ due to each systematic template with $1\sigma$ error bars.  The panel for dust is labeled ``E(B-V)'', compact sources are labeled ``Comp Sources'' with a given frequency, and the last two panels are for SZ and Cold Clumps.  It appears that the biases from compact sources and SZ are significantly deviant from null at some scales, but overall our $C_\ell^{\kappa g}$ measurement does not appear to be biased from these systematic templates.}
\end{center}
\end{figure}

We use the cross-correlations to estimate the bias to $C_\ell^{\kappa g}$ due to each systematic effect.  Assuming the formalism in \citet{2011MNRAS.417.1350R} and \citet{2012ApJ...761...14H}, where the total measured perturbation in $\kappa$ or the galaxies is a linear combination of the true perturbation and templates for the systematics, it has been shown \citep{2016MNRAS.456.3213G} that the bias and error for one systematic is given by
\begin{eqnarray}\label{E:clkgbias}
\Delta C_{\ell,i}^{\kappa g}&=&\frac{C_\ell^{\kappa M_i}C_\ell^{gM_i}}{C_\ell^{M_iM_i}}\nonumber\\
\sigma^2\left(\Delta C_{\ell,i}^{\kappa g}\right)&=&\left(\Delta C_{\ell,i}^{\kappa g}\right)^2\left[\left(\frac{\sigma(C_\ell^{\kappa M_i})}{C_\ell^{\kappa M_i}}\right)^2+\left(\frac{\sigma(C_\ell^{gM_i})}{C_\ell^{gM_i}}\right)^2\right]\, ,
\end{eqnarray}
where $i$ denotes one of the 6 dust/point source maps we consider and $C_\ell^{M_iM_i}$ is the auto-correlation for the systematic map $M_i$.  This formalism can be easily extended to find the total bias including all the systematics; however, we do not attempt this because the error on the bias becomes comparable to the magnitude of $C_\ell^{\kappa g}$.  In Fig.~\ref{F:cmbsys} we plot $\Delta C_{\ell,i}^{\kappa g}$ for each systematic.  We find that most of the bias measurements are less than 1$\sigma$ error from a null result, even more than expected from a normal distribution.  In addition, all biases are less than 2$\sigma$ error from the null result.  Therefore, we do not report from this calculation any evidence for significant bias due to any of the tested systematic templates in our $C_\ell^{\kappa g}$ measurement.

We then consider the bias from each systematic as an estimate of the bias for $C_\ell^{\kappa g}$, and we estimate the systematic error due to these contaminants by calculating the spread of the biases at each angular scale, which are then added in quadrature to estimate the full systematic error.  We define the spread in bias values as the average absolute value of $\Delta C_{\ell,i}^{\kappa g}$, weighted by $1/\sigma^2\left(\Delta C_{\ell,i}^{\kappa g}\right)$.  This procedure gives an estimate of 2.7\% for the systematic error.

% We estimate the systematic error to $E_G$ from dust and point sources by modeling the measured $C_\ell^{\kappa g}$ as
%\begin{eqnarray}\label{E:clkgbias}
%C_\ell^{\kappa g}=[C_\ell^{\kappa g}]_{\rm true}+\sum_{i=1}^6\sum_{j=1}^6\epsilon_{i,\ell}^\kappa\epsilon_{j,\ell}^gC_\ell^{M_iM_j}\, ,
%\end{eqnarray}
% and $M_j$ (or auto-correlation of $i=j$), and $\epsilon_{i,\ell}^\kappa$ and $\epsilon_{i,\ell}^g$ linearly relate the systematic perturbation to the CMB lensing and galaxy overdensity fields, respectively, in harmonic space.  This formalism has been used \citep{2011MNRAS.417.1350R,2012ApJ...761...14H} to correct for systematic effects in CMASS photometric galaxy clustering.  We solve for $\epsilon_{i,\ell}^\kappa$ and $\epsilon_{i,\ell}^g$ using the equations
%\begin{eqnarray}
%C_\ell^{XM_i}=\sum_{j=1}^6C_\ell^{M_iM_j}\epsilon_{j,\ell}^X\, ,
%\end{eqnarray}
%where $X$ is either $\kappa$ or $g$.  By solving for $\epsilon_{i,\ell}^\kappa$ and $\epsilon_{i,\ell}^g$ and inserting them into equation (\ref{E:clkgbias}), we find the bias is approximately 0.5\%.

%The cross-correlations between our observed fields (CMB lensing, galaxy overdensity) and possible systematics (dust, point sources).  CMB lensing (galaxy overdensity) cross-correlations are denoted by blue (green) crosses with $1\sigma$ error bars, with dotted crosses denoting negative values.  The panel for dust is labeled ``E(B-V)'', compact sources are labeled ``Comp Sources'' with a given frequency, and the last two panels are for SZ and Cold Clumps.  We see that compact sources and SZ are correlated with both the CMB lensing and galaxy overdensity maps, meaning they bias the CMB lensing-galaxy cross-correlation.

Redshift space distortions can also systematically reduce $E_G$ by introducing an extra correlation \citep{2007MNRAS.378..852P} in the galaxy angular power spectrum on large scales.  We find a 1.4\% effect based on the largest angular scale we use ($\ell=62$).  At smaller scales this effect's magnitude decreases, and we estimate that the effect on $E_G$ marginalized over scale is approximately 0.7\% of $E_G$.

\begin{figure}
\begin{center}
\includegraphics[width=0.5\textwidth]{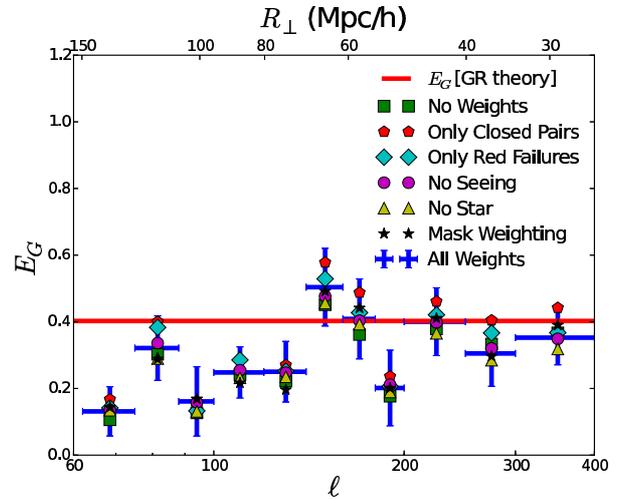}
\caption{\label{F:egsysgal} The observed $E_G$ results from mock/simulations using systematic weights [blue crosses; see Fig.~\ref{F:egv}] along with markers denoting the $E_G$ estimates with various systematic weights turned off, as well as an $E_G$ with pixel mask weighting turned on.  Most of these scenarios do not shift the $E_G$ measurement significantly.  Removing stellar and seeing weightings does shift the measurement, but not when all weights are removed.}
\end{center}
\end{figure}

We test the effects of the systematic weights for the galaxy sample by estimating $E_G$ (see section \ref{S:estim}) with various weights turned off.  We also estimate $E_G$ with pixels weighted by observed area.  Note that for the systematic weights, the shift in $E_G$ includes shifts in $C_\ell^{\kappa g}$, $C_\ell^{gg}$, and $\beta$, while for the pixel weighting we do not include a shift in $\beta$ because it is fitted from a 3D correlation function in which the completeness is already included.  In the results shown in Fig.~\ref{F:egsysgal}, we see that removing weights does not change our result by more than $1\sigma$.  We also see that weighting the pixels by the observed area (or completeness) would not shift the results significantly either.

We also consider the galaxy weights (see section \ref{S:cmbgalsys}) as a source of systematic error. The scatter in $E_G$ among all the combinations of weights we consider implies a systematic error that is $\sim11$\% of $E_G$.  However, this includes the two $E_G$ values assuming only close pair weights and only redshift failure weights, which appear to be systematically higher than the other $E_G$ values.  It has been shown \citep{2014MNRAS.441...24A} that stellar and seeing weights are necessary to produce unbiased estimates of the power spectrum, thus we will not consider the two systematically high $E_G$ values in our error estimate.  Under this assumption, the systematic error due to galaxy weights is approximately 4.5\% of $E_G$.  Note that this estimate properly combines the individual systematic errors in the angular power spectra and $\beta$ into an error for $E_G$.  This error is much greater than all the other sources of error, including from systematic corrections to $E_G$, $C_\ell^{\kappa g}$ bias, and the RSD bias in $C_\ell^{gg}$; adding all the effects together in quadrature, we find a full systematic error estimate of 5.4\%, which is much less than the statistical errors in $E_G$.

\section{Conclusions} \label{S:conclude}

$E_G$ is a bias-independent probe of gravity on large cosmological scales, and we provide the first measurement of  this quantity using CMB lensing. We construct our measurement using data from the \emph{Planck} CMB lensing map and the CMASS galaxy sample.  By using the CMB to trace the gravitational lensing field, ours is the largest-scale measurement of $E_G$ attempted.  While our measurement was not precise enough to confirm or rule out alternatives to GR, such as $f(R)$ gravity, our measurement serves as a ``first step'' towards much more precise measurements of $E_G$ from upcoming galaxy surveys, such as the Dark Energy Spectroscopic Instrument (DESI) \citep{2013arXiv1308.0847L}, the Large Synoptic Survey Telescope (LSST) \citep{2009arXiv0912.0201L}, and \emph{Euclid} \citep{2011arXiv1110.3193L} combined with next-generation CMB surveys such as Advanced ACTPol \citep{2016JLTP..tmp..144H}.

A major forthcoming challenge will be mitigating systematics in upcoming measurements.  The statistical errors in our measurements were much larger than our systematic errors, but this will not be the case with percent-level statistical errors from upcoming surveys.  Specifically, foregrounds like stellar contamination for galaxies and point sources will have to be better identified and controlled in future $E_G$ measurements.  One thing to note, however, is that $E_G$ is particularly robust to systematics in the CMB map in that those same systematics would also have to contaminate the galaxy map in order to bias $E_G$.

Finally, this work should spur future work in new probes of $E_G$.  For example, intensity mapping \citep{2010JCAP...11..016V} of 21-cm line emission will produce low-angular-resolution maps of large-scale structure (LSS).  Since $E_G$ probes gravity on large scales, high angular resolution is not necessary, allowing intensity mapping to replace the galaxy map in the $E_G$ estimator with advantages of high redshift precision and high LSS sampling.  In addition, it has been predicted \citep{2015arXiv150903286P} that the Square Kilometer Array\footnote{www.skatelescope.org} could measure the galaxy-lensing cross-correlation from intensity mapping with high precision and at multiple source redshifts.  Thus, measurements of $E_G$ using intensity mapping could serve as the supreme modified gravity probe.

\section*{Acknowledgments}
We thank R.~Mandelbaum for helpful comments on our draft.  We also thank O.~Dor\'{e}, D.~Hanson, B.~Sherwin, D.~Spergel, J.-L.~Starck, and A.~van Engelen for comments concerning the CMB lensing map, S.~Fromenteau for discussions concerning the population of galaxies in dark matter halos, S.~Singh for discussions concerning RSD bias on $E_G$, and R.~O'Connell for discussions on covariance estimates.  Finally, we thank M.~White for providing the TreePM simulations used in our analysis.  A.P. was supported by a McWilliams Fellowship of the Bruce and Astrid McWilliams Center for Cosmology.  S.A. and S.~Ho are supported by NASA grants 12-EUCLID11-0004 for this work. SH is also supported by DOE and NSF AST1412966.
 %C.~Hirata, R.~Mandelbaum, and M.~White
 
This work is based on observations obtained with \emph{Planck} (\url{http://www.esa.int/Planck}), an ESA science mission with instruments and contributions directly funded by ESA Member States, NASA, and Canada.
 
Funding for SDSS-III has been provided by the Alfred P. Sloan Foundation, the Participating Institutions, the National Science Foundation, and the U.S. Department of Energy Office of Science. The SDSS-III web site is \url{http://www.sdss3.org/}.

SDSS-III is managed by the Astrophysical Research Consortium for the Participating Institutions of the SDSS-III Collaboration including the University of Arizona, the Brazilian Participation Group, Brookhaven National Laboratory, Carnegie Mellon University, University of Florida, the French Participation Group, the German Participation Group, Harvard University, the Instituto de Astrofisica de Canarias, the Michigan State/Notre Dame/JINA Participation Group, Johns Hopkins University, Lawrence Berkeley National Laboratory, Max Planck Institute for Astrophysics, Max Planck Institute for Extraterrestrial Physics, New Mexico State University, New York University, Ohio State University, Pennsylvania State University, University of Portsmouth, Princeton University, the Spanish Participation Group, University of Tokyo, University of Utah, Vanderbilt University, University of Virginia, University of Washington, and Yale University.

\bibliographystyle{mnras}
%\bibliography{/Users/apullen/Research/mybib.bib}

\label{lastpage}
\end{document}